\documentclass[conference,letterpaper]{IEEEtran}
\IEEEoverridecommandlockouts

\usepackage{amsmath,graphicx}
\usepackage{amsfonts,xcolor}
\usepackage{flushend}

\title{Multi-Task Learning for Screen Content Image Coding}
\author{\IEEEauthorblockN{Rashid Zamanshoar Heris and Ivan V. Baji\'{c}}
\IEEEauthorblockA{\textit{School of Engineering Science, Simon Fraser University, Burnaby, BC, V5A 1S6, Canada}}
}

\usepackage{fancyhdr}
\fancypagestyle{firstpage}{
  \fancyhf{}
  \fancyhead[C]{\small To be presented at the \textit{IEEE International Symposium on Circuits and Systems (ISCAS)}, Monterey, CA, USA, May 2023.}
}

\begin{document}
%
\maketitle
\begin{abstract}
With the rise of remote work and collaboration, compression of screen content images (SCI) is becoming increasingly important. While there are efficient codecs for natural images, as well as codecs for purely-synthetic images, those SCIs that contain both synthetic and natural content pose a particular challenge. In this paper, we propose a learning-based image coding model developed for such SCIs. By training an encoder to provide a latent representation suitable for two tasks -- input reconstruction and synthetic/natural region segmentation -- we create an effective SCI image codec whose strong performance is verified through experiments. Once trained, the second task (segmentation) need not be used; the codec still benefits from the segmentation-friendly latent representation. 


\end{abstract}

\thispagestyle{firstpage}

\section{Introduction}
\label{sec:intro}

\begin{figure}[t]
 \centering
  \includegraphics[width=0.49\textwidth]{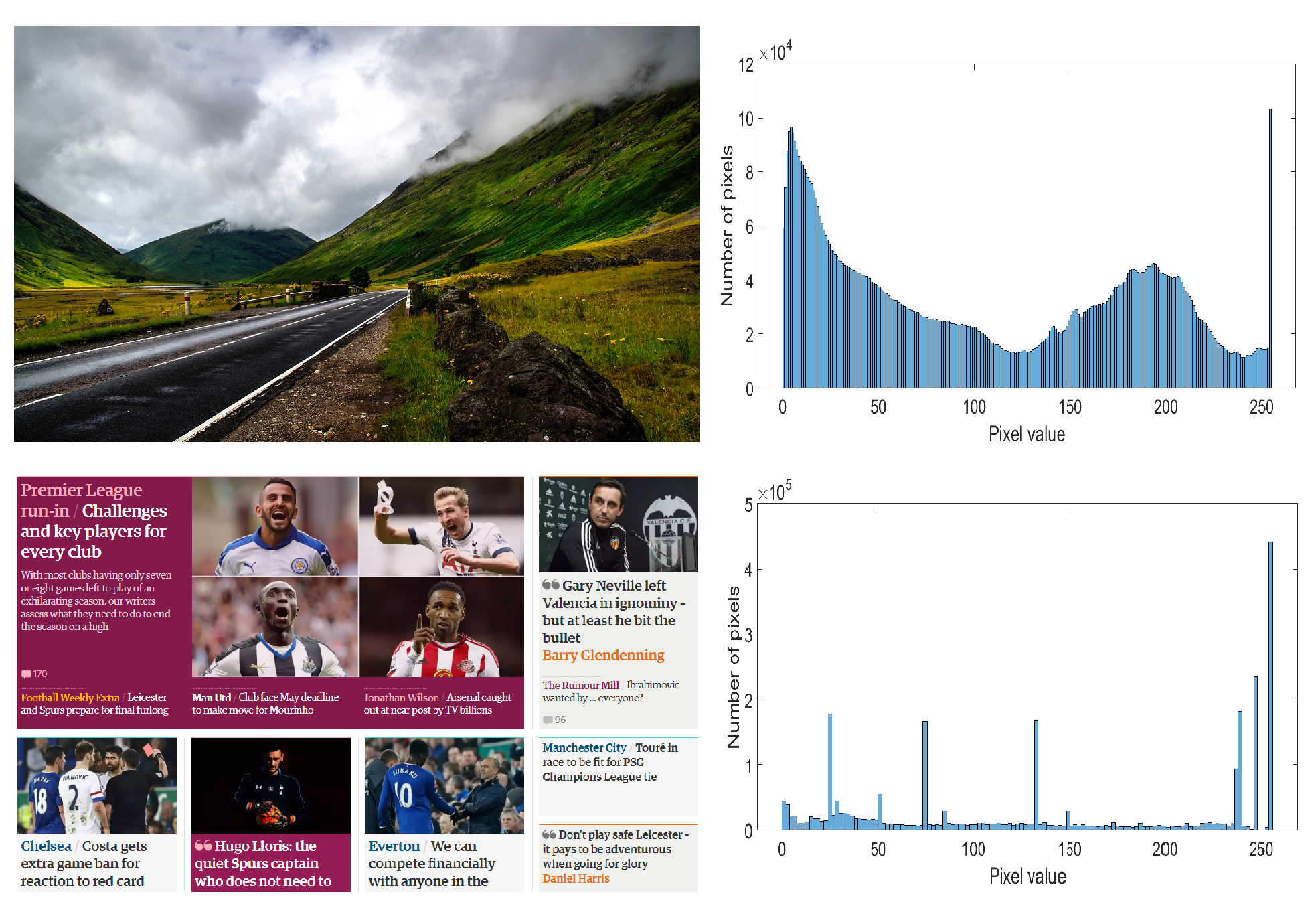}
 \caption{Top: a natural image from the CLIC dataset~\cite{CLIC2020} and its luma histogram. Bottom: a SCI from the SCID dataset~\cite{ni2017esim} and its luma histogram.}
 \label{fig:img-sample}
 \end{figure}

Traditional image compression standards such as JPEG~\cite{wallace1991jpeg} and JPEG2000~\cite{rabbani2002overview} were developed mostly with natural images in mind. However, with the rise of remote work and collaboration, transmission of screen content images (SCI) has become important. Recognizing this, a screen content coding extension~\cite{HEVC-SCC} was developed for High Efficiency Video Coding (HEVC).   
More recently, low-level coding techniques tailored to screen content 
were introduced into the Versatile Video Coding (VVC) standard~\cite{peng2016overview,bross2021overview,xu2021overview}. Due to the unique signal characteristics of SCIs, such as sharp edges, repetitive patterns, and the absence of noise, many encoding techniques that are effective for natural images are either ineffective or less efficient for SCIs~\cite{nguyen2021overview}. As an example, Fig.~\ref{fig:img-sample} shows one natural and one screen content image along with their luminance histograms. It is easy to see that the histograms 
are very different. Also the effect of coding artifacts in the two types of images can be different~\cite{SCI_Quality_wavelet}. For instance, quality loss near the edges is a common side effect of compressing natural images; yet, if done right, it may be imperceptible. However, in synthetic images, blurring of sharp edges (e.g., in SCIs showing text) can be quite noticeable and annoying, and may affect the ability of the viewer to understand the text. 
 

In the meantime, steady progress has been made in learning-based image compression~\cite{balle2016end,balle2016end2, balle2018variational,balle2018autoregressive,minnen2020channel,cheng2020learned,endtoend_img_comp}, however, the focus has been mostly   
on natural images. 
When used on SCIs, coding models trained on natural images  tend to be less effective, as will be seen in our results. Therefore, in this paper, we develop a learning-based coding model targeted at SCI. Note that SCIs can contain both synthetic and natural content. In fact, one of the key challenges in SCI compression is to adapt coding to the nature of the content (synthetic vs. natural), because these have different statistics and therefore may require different coding tools. 
We leverage the fact that a learning-based model can be trained to distinguish natural from synthetic content, as well as perform compression, in order to boost the coding performance on SCI.  

The 
paper is structured as follows. In Section~\ref{sec:r-work}, we provide a brief overview of 
SCI compression and learning-based image compression, and outline our contribution. The proposed SCI coding model 
is presented in Section~\ref{sec:proposed}.
Experiments are described in Section~\ref{sec:exp-resu} followed by conclusions in Section~\ref{sec:concl}. 

\section{Related work and our contribution}
\label{sec:r-work}

\subsection{Traditional screen content image compression}
\label{ssec:r-work-sc}
High-Efficiency Video Coding (HEVC) standard contains specialized tools for screen content compression, which have been wrapped into its screen content coding extension (HEVC-SCC)~\cite{HEVC-SCC,peng2016overview}. 
These include pallete mode~\cite{pu2016palette}, sample-based prediction
~\cite{kang2016efficient}, and DPCM-based edge prediction
~\cite{sanchez2016dpcm}. Line-by-line template matching, which is more 
flexible than HEVC's native intra-block-copy mode, has been proposed in~\cite{peng2016hash}. In \cite{mitrica2019very}, a layer-based method is presented for extracting the synthetic data from the screen content image. While the remaining natural video is compressed using HEVC-SCC, detected text and graphical structures are encoded using a customized compression strategy. In~\cite{wang2016utility}, a pre-processing step is suggested 
to alter parts of the original image data based on the utility of the related image region in order to decrease the bitrate. 
Additionally,~\cite{yang2015perceptual} and~\cite{gu2016saliency} have examined the quality of screen content images when using lossy compression. 
Versatile Video Coding (VVC) standard~\cite{nguyen2021overview} also includes 
low-level coding tools that are suitable for screen content coding. 

\subsection{Learning-based image compression}
\label{ssec:r-work-dl}
\begin{figure}[t]
 \centering
  \includegraphics[width=0.48\textwidth]{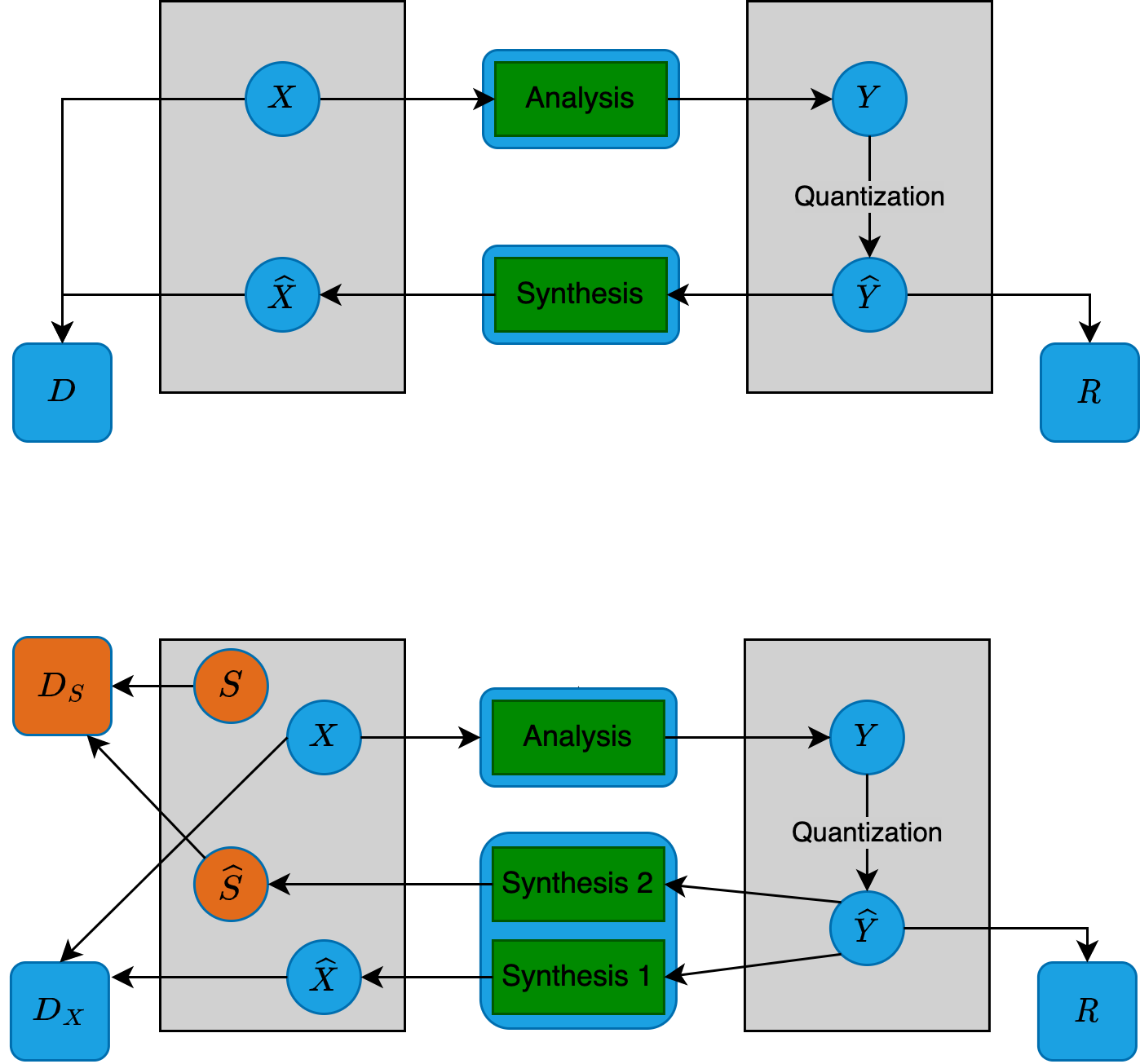}
 \caption{Top: high-level structure of a learning-based image codec. Bottom: proposed approach.}
 \label{fig:learning-based-codecs}
\end{figure}

Learning-based image compression has made rapid progress in recent years. Early works~\cite{toderici-variable, Minnen_ICIP2017,Johnston_CVPR2018} employed Recurrent Neural Networks (RNNs) to model spatial relationships in an image. More recent models are based Convolutional Neural Networks (CNNs) and have a structure shown in the top part of Fig.~\ref{fig:learning-based-codecs}. An image $X$ is subject to a nonlinear analysis transform~\cite{balle2016end,balle2016end2}, composed of convolutional layers, nonlinear activations, and normalizations, among which generalized divisive normalization (GDN)~\cite{balle2015density} is a popular choice. The latent representation $Y$ is then quantized to $\widehat{Y}$ and entropy-coded to produce the compressed bit stream. An important part of the learning process is estimating the rate $R$, i.e., modeling the entropy of $\widehat{Y}$. Entropy models have become more sophisticated in recent years, involving factorized priors and hyperpriors~\cite{balle2018variational}, autoregressive and hierarchical priors~\cite{balle2018autoregressive}, channel-wise autoregressive models~\cite{minnen2020channel}, and Gaussian mixtures~\cite{cheng2020learned}. Most recent work has explored advanced learned transforms such as normalizing flows~\cite{ANFIC} and transformers~\cite{VR_Transformer}. Common to all end-to-end learning-based codecs is that differentiable estimates of rate $R$ and distortion $D$ are use to optimize the whole system~\cite{endtoend_img_comp}.

Learning-based image compression has largely focused on natural images so far. However, recently,~\cite{transform_skip_ICIP22} presented a learned image codec that incorporates the concept of transform skip into the end-to-end pipeline to improve SCI compression. The authors also showed that retraining an end-to-end learned pipeline on SCI would improve SCI compression performance, indicating the effect of ``domain shift'' between natural and screen content images. Our paper provides further quantification of the effects of this domain shift, and proposes an alternative training strategy to improve SCI coding performance.

\subsection{Our contribution}
Our contributions can be summarized as follows:
\begin{itemize}
    \item We quantify the performance gap of two learning-based codecs~\cite{balle2018variational,minnen2020channel} due to ``domain shift'' from natural to synthetic and mixed natural-synthetic content.
    \item We develop a multi-task SCI coding approach that is able to distinguish natural from synthetic content in SCI, to help boost coding performance.
    \item We demonstrate the strong coding performance of the proposed approach by comparing it against relevant benchmarks, both handcrafted and learned.  
\end{itemize}

\begin{figure}[t] 
 \centering  \includegraphics[width=0.45\textwidth,height=0.17\textheight]{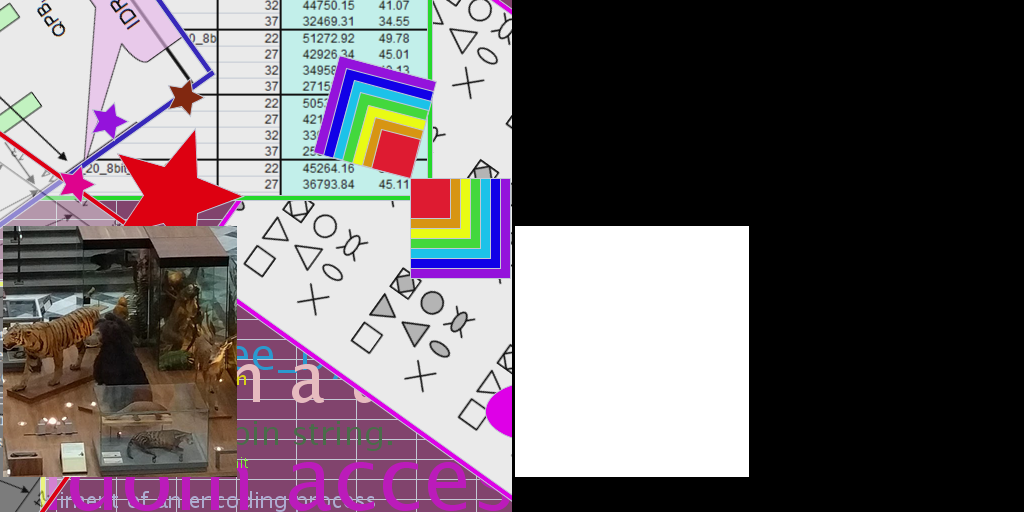}
 \caption{An example of a SCI used in our experiments (left), with its ground-truth synthetic/natural (S/N) segmentation map (right).}
 \label{fig:SC-N}
\end{figure}

\section{Proposed methods
}
\label{sec:proposed}

\subsection{Multi-task coding model}
The main idea behind the proposed SCI codec is that we want to create a latent representation that is both compressible and has sufficient information about synthetic and natural parts of the input image, to help the codec adjust to different statistics of these regions. To accomplish this, we propose a multi-task architecture shown in the bottom of Fig.~\ref{fig:learning-based-codecs}. Image $X$ is input to learnable analysis transform $T_a$ to produce the latent representation $Y$ which is then quantized into $\widehat{Y}$:
\begin{equation}
    Y = T_a(X), \qquad \widehat{Y} = Q(Y).
\end{equation}
The quantized representation $\widehat{Y}$ is entropy coded to produce the compressed bit stream. At the decoder, the bit stream is entropy decoded to recover $\widehat{Y}$, which is then passed to two synthesis transforms,
\begin{equation}
    \widehat{X} = T_s^{(1)}(\widehat{Y}), \qquad \widehat{S} = T_s^{(2)}(\widehat{Y}),
\end{equation}
to generate the reconstructed input image $\widehat{X}$ and the synthetic/natural (S/N) segmentation map $\widehat{S}$, which is supposed to separate synthetic from natural parts of the SCI image $X$. An example of such a segmentation map is shown in Fig.~\ref{fig:SC-N} next to its corresponding input image. We describe later in this section how we generate input data (both SCIs and S/N segmentation maps) for training the proposed system.

Following~\cite{balle2018variational}, the rate is estimated as
\begin{equation}
R=\mathbb{E}\left [ -\textup{log}_{2}p_{\widehat{Y}}(\widehat{Y}) \right ] + \mathbb{E}\left [ -\textup{log}_{2}p_{\widehat{Z}}(\widehat{Z}) \right ],
\label{eq:rate_term}
\end{equation}
where $\widehat{Z}$ is the quantized hyperprior used in entropy modeling, $p_{\widehat{Y}}$ and $p_{\widehat{Z}}$ are the probability distributions of $\widehat{Y}$ and $\widehat{Z}$, respectively, and the expectation is taken over input data. The loss function used in the training is
\begin{equation}
    \mathcal{L} = R + \lambda \cdot D_X + \varphi \cdot D_S,
    \label{eq:loss}
\end{equation}
where $D_X$ and $D_S$ are distortion measures for input reconstruction and S/N segmentation, respectively. 
We used $D_X = \text{MSE}(X,\widehat{X})$ and $D_S = \text{MSE}(S,\widehat{S})$ in the experiments, but other choices are possible, 
such as 
SSIM~\cite{wang2004image} for reconstruction or Intersection-over-Union for segmentation. Once the model is trained, the segmentation output need not be used; it's purpose was to 
make the system aware of synthetic/natural parts of the input.

Note that any transforms, quantizers, and entropy models from learning-based coding pipelines can be used to construct the proposed multi-task codec. To illustrate this, we conducted our experiments with components from two pipelines, namely~\cite{balle2018variational} and~\cite{minnen2020channel}, and show that in both cases, the proposed multi-task approach improves SCI coding performance compared to the default pipeline.  

\vspace{-3pt}
\subsection{Dataset}
\label{sec:train-p}
The proposed multi-task model requires ground-truth S/N segmentation masks for training, to be able to compute $D_S$. Hence, we constructed such a dataset\footnote{\texttt{https://github.com/SFU-Multimedia-Lab/MLSCIC}} from synthetic and natural images. For this purpose, we used synthetic SCIs from~\cite{ni2017esim} and~\cite{cheng2020screen}, and natural images from~\cite{CLIC2020}. The dataset is constructed as follows. First, a pair of synthetic and natural images is randomly chosen. Then, a random patch (with randomly chosen size, such that width and height are between 128 and 192 pixels) is extracted from the natural image 
and inserted into a randomly chosen location in the synthetic image. At the same time, the size and location of the natural patch within the synthetic image are used to construct the binary S/N segmentation map, as shown in Fig.~\ref{fig:SC-N}. This way, a dataset of 3,100 images and their corresponding S/N segmentation masks is constructed, both with size 512 $\times$ 512. Of these, 100 randomly chosen images are used for testing and the remaining 3,000 images are used for training. 

\section{Experiments}
\label{sec:exp-resu}
\vspace{-7pt}
\subsection{Setup}
We construct the multi-task codec using two learning-based image compression pipelines --~\cite{balle2018variational} and~\cite{minnen2020channel} -- referred to as ``bmshj2018'' and ``ms2020'', respectively. In each case, another synthesis transform block, architecturally the exact copy of the existing synthesis block, is added to the decoder to provide the S/N segmentation output. The only difference is that the output layer of the added synthesis block is changed to produce a 1-channel (binary) image as the segmentation map, instead of the 3-channel RGB image.

Both default pipelines are trained on the CLIC dataset~\cite{CLIC2020} for 1,000 epochs using the Adam optimizer~\cite{kingma2014adam} with an initial learning rate of $10^{-4}$, to provide a baseline trained on natural images. 
These are called X-1-Decoder-Natural, where X $\in$ \{``bmshj2018'', ``ms2020''\}. Further, the same pipelines are trained in the same way on synthetic images from~\cite{ni2017esim} and~\cite{cheng2020screen} to provide codecs called X-1-Decoder-Synthetic,  
and on our synthetic/natural dataset for 300 epochs to provide codecs called X-1-Decoder-S/N. 
Finally, our multi-task codecs are  trained on our dataset for 300 epochs, and are called X-2-Decoder. The following $(\lambda,\varphi)$ pairs were used in~(\ref{eq:loss}) to train five \, multi-task \, codecs: \, $(\lambda,\varphi) \, \in \, \{(1, 10^{-5}), \, (10^{-1}, 10^{-5}), \\ (10^{-2}, 10^{-6}), (10^{-3}, 10^{-7}), (10^{-4}, 10^{-7})\}$. The single-task codecs used only the $\lambda$ values from this list, without the last term in the loss function~(\ref{eq:loss}).
Two image quality metrics are used in the experiments: 
Peak Signal to Noise Ratio (PSNR), and Gradient Magnitude Similarity Deviation (GMSD)~\cite{xue2013gradient}, which is considered suitable for SCIs.


\begin{table}[tb]
    \centering
    \caption{BD-Rates (\%) for PSNR and GMSD metrics relative to bmshj2018-1-Decoder-Natural for bmshj2018-based codecs and ms2020-1-Decoder-Natural for ms2020-based codecs} 
    \label{tab:BD-Rate-bmshj-ms}
    \begin{tabular}{|c|r|r|}
    \hline
        Codec & PSNR &  GMSD  \\
        \hline \hline \hline 
        bmshj2018-1-Decoder-Synthetic   & --20.45& --20.83 \\ \hline
        bmshj2018-1-Decoder-S/N         & --25.87 & --31.08  \\ \hline
        bmshj2018-2-Decoder             & \textbf{--36.14} &  \textbf{--34.48} \\ \hline \hline
        ms2020-1-Decoder-Synthetic     & --17.39 &  0.15 \\ \hline    
        ms2020-1-Decoder-S/N    & --39.97  &  --43.68\\ \hline
        ms2020-2-Decoder               & \textbf{--61.57} &  \textbf{--70.95} \\ \hline
        
    \end{tabular}
    \vspace{-10pt}
\end{table}

\vspace{-3pt}
\subsection{Quantifying the domain shift}
Table~\ref{tab:BD-Rate-bmshj-ms} quantifies various domain shifts for the two coding pipelines in terms of the BR-Rate~\cite{bjontegaard2001calculation}, for both PSNR and GMSD. For each pipeline, X-1-Decoder-Natural, where X $\in$ \{``bmshj2018'', ``ms2020''\}, is used as the anchor, and best results for each pipeline are indicated in bold. The results show that merely retraining a coding model on synthetic images will reduce the BD-Rate by 17-20\% when the quality metric is PSNR. This provides additional support to the results from~\cite{transform_skip_ICIP22}, which showed similar trend. 
Moreover, our results show that retraining the pipeline on combined synthetic/natural images provides 20-40\% BD-Rate gain for PSNR. Finally, the multi-task system provides 36-62\% BD-Rate gain for PSNR. 

The comparison between X-1-Decoder-S/N and X-2-Decoder is particularly interesting. Both these types of codecs are trained our synthetic/natural dataset, and the only difference is that X-2-Decoder is performing two tasks instead of one. Hence, this comparison shows the effect of multi-tasking. We see from Table~\ref{tab:BD-Rate-bmshj-ms} that adding the second task (S/N segmentation) improves BD-Rate by 10-20\% for PSNR, and 3-35\% for GMSD, depending on the pipeline X. We hypothesize that this is because the encoder in a X-2-Decoder system is more aware of where natural vs. synthetic parts of the input are, so it is able to adjust its entropy model parameters more quickly to the different statistics of these regions. 

\subsection{Screen content image coding performance}
Next we test the SCI image coding performance on our test set. Since the ``ms2020'' pipeline showed better performance in the experiments described above, here we use ms2020-1-Decoder-Natural as the anchor, and examine the coding performance of other versions of this pipeline. Further, we also test HEVC Intra (HM 16.26rc1), HEVC-SCC Intra, and VVC Intra (VTM 18.2) with screen content coding options turned on.  
Table~\ref{tab:BD-Rate-ms} shows the BD-Rate relative to the ms2020-1-Decoder-Natural anchor. As seen in the table, among the conventional codecs, VVC is the best with about 46\% BD rate gain over the anchor for both PSNR and GMSD. Both HEVC-SCC and ms2020-1-Decoder-S/N are close behind it for GMSD, with a slightly larger gap for PSNR. However, our ms2020-2-Decoder provides the best performance, with over 60\% (resp. 70\%) BD-Rate gain for PSNR (resp. GMSD).


  
\begin{table}[tb]
    \centering
    \caption{BD-Rates (\%) for PSNR and GMSD metrics relative to ms2020-1-Decoder-Natural}
    \label{tab:BD-Rate-ms}
    \begin{tabular}{|c|r|r|}
    \hline
        Codec & PSNR &  GMSD  \\
        \hline \hline \hline 
        HEVC (HM 16.26rc1) & --10.37  & --28.31 \\ \hline
        HEVC-SCC    &  --37.61 & --43.92 \\ \hline
        VVC (VTM 18.2) & --46.31  & --46.02 \\ \hline
        ms2020-1-Decoder-Synthetic    &  --17.39 & 0.15 \\ \hline
        ms2020-1-Decoder-S/N & --39.97  & --43.68 \\ \hline
        ms2020-2-Decoder    &  \textbf{--61.57} & \textbf{--70.95} \\ \hline
    \end{tabular}
\end{table}

\begin{figure}[tb]
    \centering
    \includegraphics[width=0.49\columnwidth, trim={0 4cm 0 0},clip]{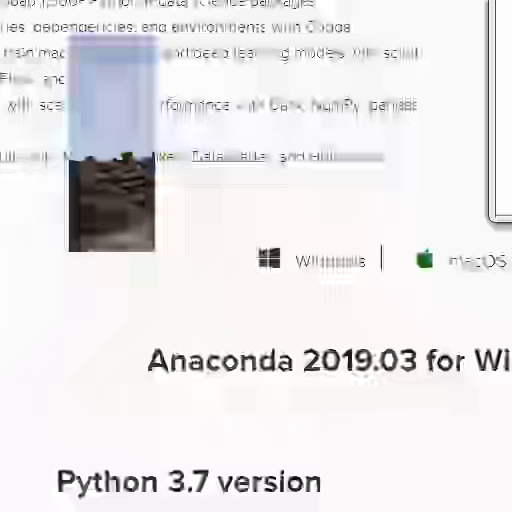}
    \includegraphics[width=0.49\columnwidth, trim={0 4cm 0 0},clip]{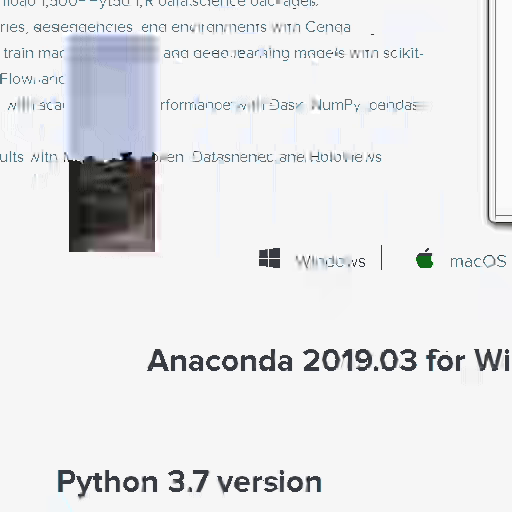}
    \includegraphics[width=0.49\columnwidth, trim={0 4cm 0 0},clip]{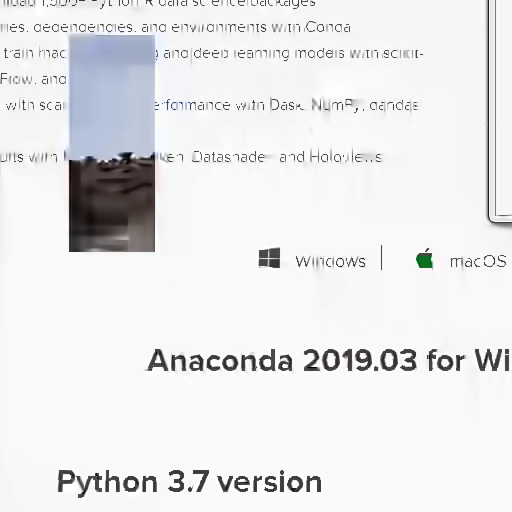}
    \includegraphics[width=0.49\columnwidth, trim={0 4cm 0 0},clip]{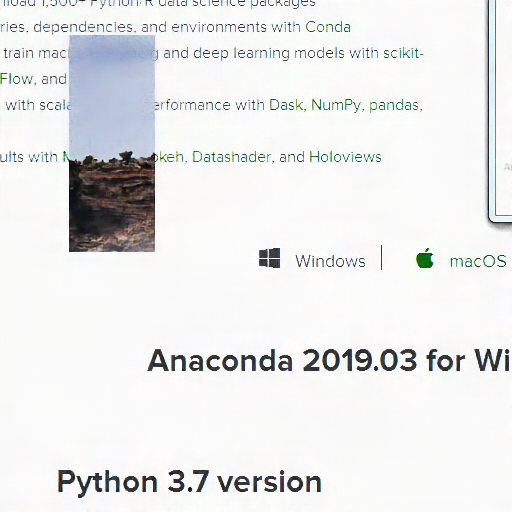}
    \caption{Visual example of SCI coding. Top-left: HEVC (0.0604 bpp, PSNR = 25.81 dB, GMSD = 0.1593). Top-right: HEVC-SCC (0.0581 bpp, PSNR = 27.84 dB, GMSD = 0.1344). Bottom-left: VVC (0.0569 bpp, PSNR = 28.13 dB, GMSD = 0.1192). Bottom-right: proposed ms2020-2-Decoder (0.0380 bpp, PSNR = 29.12 dB, GMSD = 0.1019). With GMSD, the lower the better.}
    \label{fig:visual_examples_coding}
\end{figure}

Fig.~\ref{fig:visual_examples_coding} shows an example of a SCI coded 
with HEVC, HEVC-SCC, VVC, and the proposed codec. From HEVC (shown in the top-right) to the proposed (bottom-left), the bitrate decreases but quality increases. All codecs handle large text reasonably well. However, details of both the small text and the natural part of the SCI are preserved much better with the proposed codec compared to the other codecs.

In order to further explain the effectiveness of the proposed multi-task approach, we measure its coding performance on the natural and synthetic parts of the test SCI images. Specifically, we measure the bit rate and PSNR separately on these two portions of the input image, and then compute BD-Rate with respect to the anchor ms2020-1-Decoder-Natural. The results are shown in 
Table \ref{tab:BD-Rate-ms-N-SC}. Recall that ms2020-1-Decoder-Synthetic is trained on the pure synthetic training set, and ms2020-2-Decoder is trained on our synthetic/natural training set. The results show that ms2020-1-Decoder-Synthetic has lost about 118\% BD-Rate on the natural portion compared to the anchor, but has gained slightly (about 17\% BD-Rate) on the synthetic portion compared to the anchor. Hence, improved coding of synthetic portion has been paid for by the lower efficiency on the natural portion of the SCI. However, the proposed multi-task system loses only about 26\% BD-Rate compared to the anchor on the natural portion while gaining about 69\% on the synthetic portion. Hence, in the case of the multi-task codec, the large coding gain on the synthetic portion is  paid for by a much smaller loss on the natural portion, because the codec is better able to recognize the two regions and adjust its coding accordingly.  

\begin{figure}
    \centering
    \includegraphics[width=0.49\columnwidth]{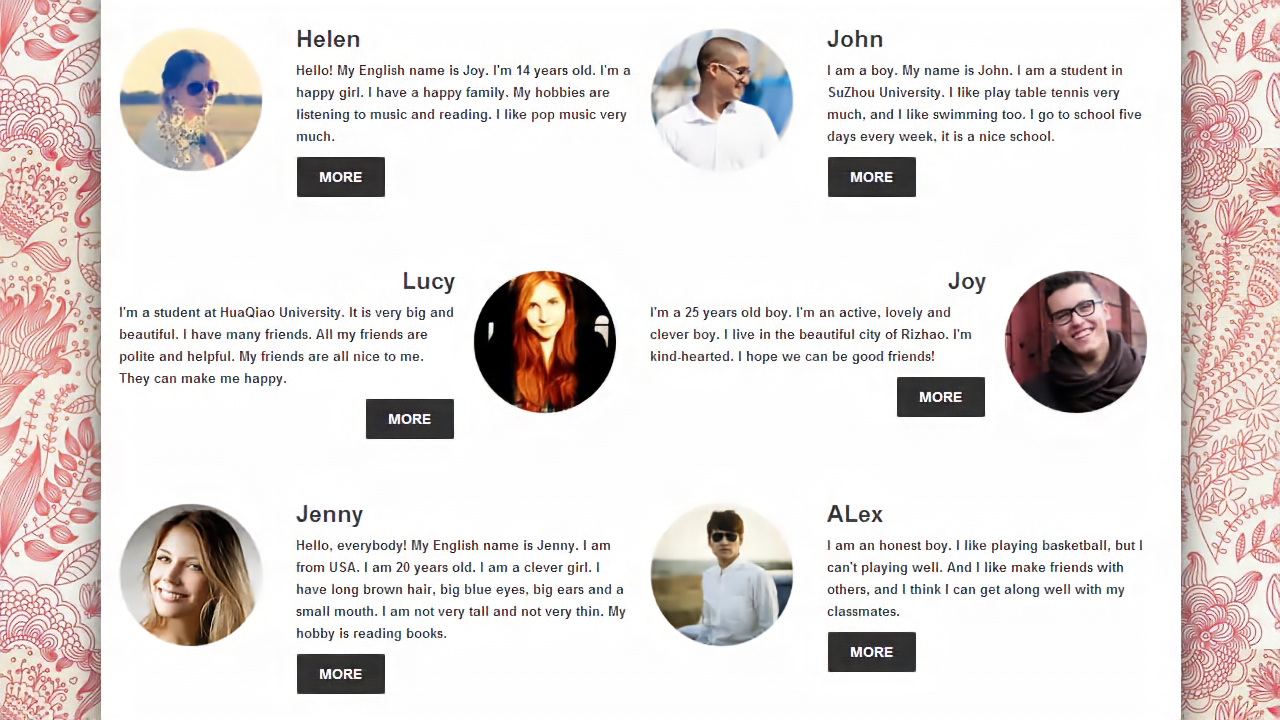}
    \includegraphics[width=0.49\columnwidth]{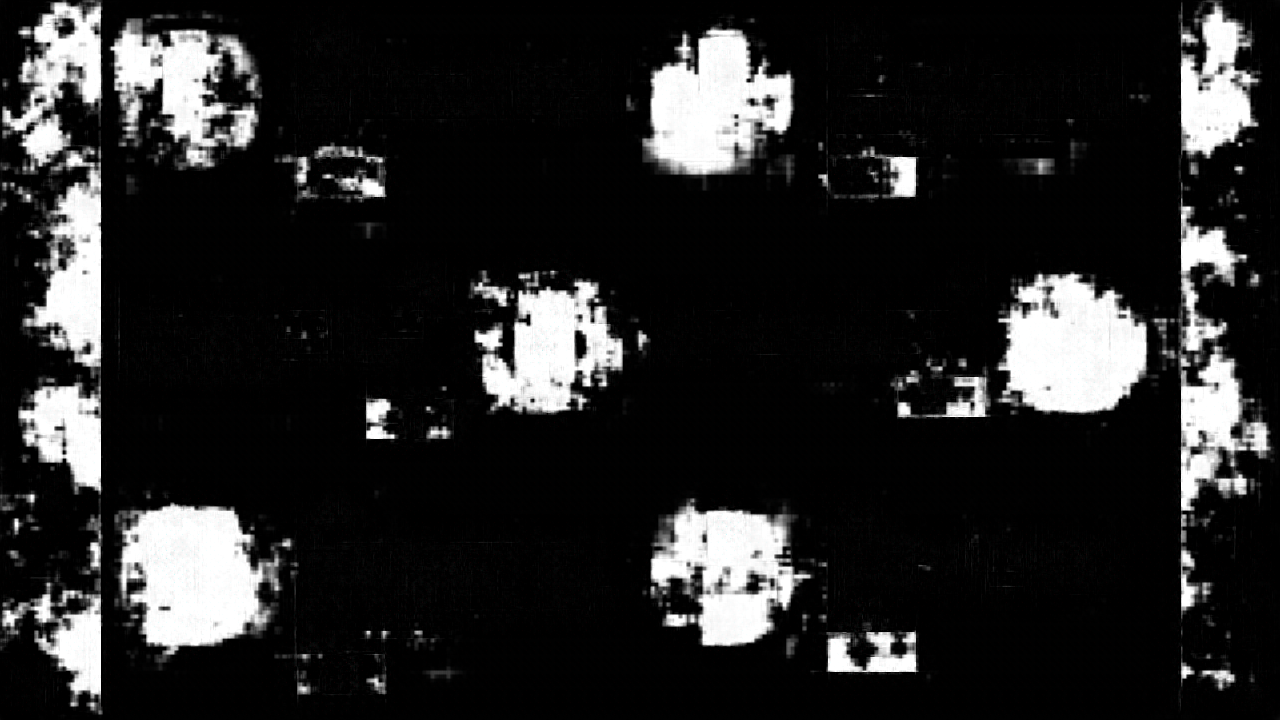}
    \caption{Example of segmentation for a SCI with multiple circular natural components.}
    \label{fig:segmentation_example}
    \vspace{-5pt}
\end{figure}

Although segmentation is not the main goal of the proposed codec, it is interesting to see how the system performs on this task as well. Fig.~\ref{fig:segmentation_example} shows a SCI that was not in our dataset, and has multiple circular natural inserts. Our multi-task system is still able to identify these natural portions reasonably well. 


\begin{table}[tb]
    \centering
    \caption{BD-Rates (\%) for PSNR in natural and synthetic parts of SCI relative to ms2020-1-Decoder-Natural}
    \label{tab:BD-Rate-ms-N-SC}
    \begin{tabular}{|c|r|r|}
    \hline
        Codec & Natural & Synthetic \\
        \hline \hline
        ms2020-1-Decoder-Synthetic & 117.63  & --16.66  \\ \hline
        ms2020-2-Decoder    &  25.61 & --68.67  \\ \hline
    \end{tabular}
\end{table}

\section{Conclusion}
\label{sec:concl}

We presented a learning-based multi-task codec for screen content images (SCI). Besides coding and reconstructing the input image, the multi-task codec was trained to perform an additional task: segment synthetic vs. natural portions of the input image. This helps it adjust its coding to the two regions, which have different statistics. The performance of the proposed codec was compared against relevant benchmarks, both handcrafted and learning-based, and its superior SCI coding performance was demonstrated through extensive experiments.


\bibliographystyle{IEEEbib}
\bibliography{strings}

\begin{thebibliography}{10}

\bibitem{wallace1991jpeg}
Gregory~K Wallace,
\newblock ``The {JPEG} still picture compression standard,''
\newblock {\em Communications of the ACM}, vol. 34, no. 4, pp. 30--44, 1991.

\bibitem{rabbani2002overview}
Majid Rabbani and Rajan Joshi,
\newblock ``An overview of the {JPEG 2000} still image compression standard,''
\newblock {\em Signal Processing: Image Communication}, vol. 17, no. 1, pp.
  3--48, 2002.

\bibitem{HEVC-SCC}
Jizheng Xu, Rajan Joshi, and Robert~A. Cohen,
\newblock ``Overview of the emerging hevc screen content coding extension,''
\newblock {\em IEEE Transactions on Circuits and Systems for Video Technology},
  vol. 26, no. 1, pp. 50--62, 2016.

\bibitem{peng2016overview}
Wen-Hsiao Peng, Frederick~G Walls, Robert~A Cohen, Jizheng Xu, J{\"o}rn
  Ostermann, Alexander MacInnis, and Tao Lin,
\newblock ``Overview of screen content video coding: Technologies, standards,
  and beyond,''
\newblock {\em IEEE Journal on Emerging and Selected Topics in Circuits and
  Systems}, vol. 6, no. 4, pp. 393--408, 2016.

\bibitem{bross2021overview}
Benjamin Bross, Ye-Kui Wang, Yan Ye, Shan Liu, Jianle Chen, Gary~J Sullivan,
  and Jens-Rainer Ohm,
\newblock ``Overview of the versatile video coding {(VVC)} standard and its
  applications,''
\newblock {\em IEEE Transactions on Circuits and Systems for Video Technology},
  vol. 31, no. 10, pp. 3736--3764, 2021.

\bibitem{xu2021overview}
Xiaozhong Xu and Shan Liu,
\newblock ``Overview of screen content coding in recently developed video
  coding standards,''
\newblock {\em IEEE Transactions on Circuits and Systems for Video Technology},
  vol. 32, no. 2, pp. 839--852, 2021.

\bibitem{nguyen2021overview}
Tung Nguyen, Xiaozhong Xu, Felix Henry, Ru-Ling Liao, Mohammed~Golam Sarwer,
  Marta Karczewicz, Yung-Hsuan Chao, Jizheng Xu, Shan Liu, Detlev Marpe, and
  Gary~J. Sullivan,
\newblock ``Overview of the screen content support in {VVC}: Applications,
  coding tools, and performance,''
\newblock {\em IEEE Transactions on Circuits and Systems for Video Technology},
  vol. 31, no. 10, pp. 3801--3817, 2021.

\bibitem{SCI_Quality_wavelet}
Pooryaa Cheraaqee, Zahra Maviz, Azadeh Mansouri, and Ahmad Mahmoudi-Aznaveh,
\newblock ``Quality assessment of screen content images in wavelet domain,''
\newblock {\em IEEE Transactions on Circuits and Systems for Video Technology},
  vol. 32, no. 2, pp. 566--578, 2022.

\bibitem{CLIC2020}
George Toderici, Wenzhe Shi, Radu Timofte, Lucas Theis, Johannes Balle, Eirikur
  Agustsson, Nick Johnston, and Fabian Mentzer,
\newblock ``Workshop and challenge on learned image compression (clic2020),''
  2020.

\bibitem{ni2017esim}
Zhangkai Ni, Lin Ma, Huanqiang Zeng, Jing Chen, Canhui Cai, and Kai-Kuang Ma,
\newblock ``Esim: Edge similarity for screen content image quality
  assessment,''
\newblock {\em IEEE Transactions on Image Processing}, vol. 26, no. 10, pp.
  4818--4831, 2017.

\bibitem{balle2016end}
Johannes Ball{\'e}, Valero Laparra, and Eero~P Simoncelli,
\newblock ``End-to-end optimization of nonlinear transform codes for perceptual
  quality,''
\newblock in {\em 2016 Picture Coding Symposium (PCS)}. IEEE, 2016, pp. 1--5.

\bibitem{balle2016end2}
Johannes Ball{\'e}, Valero Laparra, and Eero~P Simoncelli,
\newblock ``End-to-end optimized image compression,''
\newblock in {\em ICLR}, 2017.

\bibitem{balle2018variational}
Johannes Ball{\'e}, David Minnen, Saurabh Singh, Sung~Jin Hwang, and Nick
  Johnston,
\newblock ``Variational image compression with a scale hyperprior,''
\newblock in {\em ICLR}, 2018.

\bibitem{balle2018autoregressive}
Johannes Ball{\'e}, David Minnen, and George Toderici,
\newblock ``Joint autoregressive and hierarchical priors for learned image
  compression,''
\newblock in {\em NeurIPS}, 2018, vol.~31, pp. 10794--10803.

\bibitem{minnen2020channel}
David Minnen and Saurabh Singh,
\newblock ``Channel-wise autoregressive entropy models for learned image
  compression,''
\newblock in {\em 2020 IEEE International Conference on Image Processing
  (ICIP)}. IEEE, 2020, pp. 3339--3343.

\bibitem{cheng2020learned}
Zhengxue Cheng, Heming Sun, Masaru Takeuchi, and Jiro Katto,
\newblock ``Learned image compression with discretized gaussian mixture
  likelihoods and attention modules,''
\newblock in {\em Proceedings of the IEEE/CVF Conference on Computer Vision and
  Pattern Recognition}, 2020, pp. 7939--7948.

\bibitem{endtoend_img_comp}
Yueyu Hu, Wenhan Yang, Zhan Ma, and Jiaying Liu,
\newblock ``Learning end-to-end lossy image compression: A benchmark,''
\newblock {\em IEEE Transactions on Pattern Analysis and Machine Intelligence},
  vol. 44, no. 8, pp. 4194--4211, 2022.

\bibitem{pu2016palette}
Wei Pu, Marta Karczewicz, Rajan Joshi, Vadim Seregin, Feng Zou, Joel Sole,
  Yu-Chen Sun, Tzu-Der Chuang, Polin Lai, Shan Liu, et~al.,
\newblock ``Palette mode coding in {HEVC} screen content coding extension,''
\newblock {\em IEEE Journal on Emerging and Selected Topics in Circuits and
  Systems}, vol. 6, no. 4, pp. 420--432, 2016.

\bibitem{kang2016efficient}
Je-Won Kang, Soo-Kyung Ryu, Na-Young Kim, and Min-Joo Kang,
\newblock ``Efficient residual dpcm using an $ l\_1 $ robust linear prediction
  in screen content video coding,''
\newblock {\em IEEE Transactions on Multimedia}, vol. 18, no. 10, pp.
  2054--2065, 2016.

\bibitem{sanchez2016dpcm}
Victor Sanchez, Francesc Auli-Llinas, and Joan Serra-Sagrista,
\newblock ``Dpcm-based edge prediction for lossless screen content coding in
  hevc,''
\newblock {\em IEEE Journal on Emerging and Selected Topics in Circuits and
  Systems}, vol. 6, no. 4, pp. 497--507, 2016.

\bibitem{peng2016hash}
Xiulian Peng and Jizheng Xu,
\newblock ``Hash-based line-by-line template matching for lossless screen image
  coding,''
\newblock {\em IEEE Transactions on Image Processing}, vol. 25, no. 12, pp.
  5601--5609, 2016.

\bibitem{mitrica2019very}
Iulia Mitrica, Eric Mercier, Christophe Ruellan, Attilio Fiandrotti, Marco
  Cagnazzo, and B{\'e}atrice Pesquet-Popescu,
\newblock ``Very low bitrate semantic compression of airplane cockpit screen
  content,''
\newblock {\em IEEE Transactions on Multimedia}, vol. 21, no. 9, pp.
  2157--2170, 2019.

\bibitem{wang2016utility}
Shiqi Wang, Xinfeng Zhang, Xianming Liu, Jian Zhang, Siwei Ma, and Wen Gao,
\newblock ``Utility-driven adaptive preprocessing for screen content video
  compression,''
\newblock {\em IEEE Transactions on Multimedia}, vol. 19, no. 3, pp. 660--667,
  2016.

\bibitem{yang2015perceptual}
Huan Yang, Yuming Fang, and Weisi Lin,
\newblock ``Perceptual quality assessment of screen content images,''
\newblock {\em IEEE Transactions on Image Processing}, vol. 24, no. 11, pp.
  4408--4421, 2015.

\bibitem{gu2016saliency}
Ke~Gu, Shiqi Wang, Huan Yang, Weisi Lin, Guangtao Zhai, Xiaokang Yang, and
  Wenjun Zhang,
\newblock ``Saliency-guided quality assessment of screen content images,''
\newblock {\em IEEE Transactions on Multimedia}, vol. 18, no. 6, pp.
  1098--1110, 2016.

\bibitem{toderici-variable}
George Toderici, Sean~M O'Malley, Sung~Jin Hwang, Damien Vincent, David Minnen,
  Shumeet Baluja, Michele Covell, and Rahul Sukthankar,
\newblock ``Variable rate image compression with recurrent neural networks,''
\newblock in {\em ICLR}, 2016.

\bibitem{Minnen_ICIP2017}
David Minnen, George Toderici, Michele Covell, Troy Chinen, Nick Johnston, Joel
  Shor, Sung~Jin Hwang, Damien Vincent, and Saurabh Singh,
\newblock ``Spatially adaptive image compression using a tiled deep network,''
\newblock in {\em 2017 IEEE International Conference on Image Processing
  (ICIP)}, 2017, pp. 2796--2800.

\bibitem{Johnston_CVPR2018}
Nick Johnston, Damien Vincent, David Minnen, Michele Covell, Saurabh Singh,
  Troy Chinen, Sung Jin~Hwang, Joel Shor, and George Toderici,
\newblock ``Improved lossy image compression with priming and spatially
  adaptive bit rates for recurrent networks,''
\newblock in {\em Proc. IEEE/CVF CVPR}, 2018, pp. 4385--4393.

\bibitem{balle2015density}
Johannes Ball{\'e}, Valero Laparra, and Eero~P Simoncelli,
\newblock ``Density modeling of images using a generalized normalization
  transformation,''
\newblock in {\em ICLR}, 2016.

\bibitem{ANFIC}
Yung-Han Ho, Chih-Chun Chan, Wen-Hsiao Peng, Hsueh-Ming Hang, and Marek
  Domański,
\newblock ``Anfic: Image compression using augmented normalizing flows,''
\newblock {\em IEEE Open Journal of Circuits and Systems}, vol. 2, pp.
  613--626, 2021.

\bibitem{VR_Transformer}
Binglin Li, Jie Liang, and Jingning Han,
\newblock ``Variable-rate deep image compression with vision transformers,''
\newblock {\em IEEE Access}, vol. 10, pp. 50323--50334, 2022.

\bibitem{transform_skip_ICIP22}
Meng Wang, Kai Zhang, Li~Zhang, Yaojun Wu, Yue Li, Junru Li, and Shiqi Wang,
\newblock ``Transform skip inspired end-to-end compression for screen content
  image,''
\newblock in {\em Proc. IEEE ICIP}, 2022, pp. 3848--3852.

\bibitem{wang2004image}
Zhou Wang, Alan~C Bovik, Hamid~R Sheikh, and Eero~P Simoncelli,
\newblock ``Image quality assessment: from error visibility to structural
  similarity,''
\newblock {\em IEEE Transactions on Image Processing}, vol. 13, no. 4, pp.
  600--612, 2004.

\bibitem{cheng2020screen}
Shan Cheng, Huanqiang Zeng, Jing Chen, Junhui Hou, Jianqing Zhu, and Kai-Kuang
  Ma,
\newblock ``Screen content video quality assessment: Subjective and objective
  study,''
\newblock {\em IEEE Transactions on Image Processing}, vol. 29, pp. 8636--8651,
  2020.

\bibitem{kingma2014adam}
Diederik~P Kingma and Jimmy Ba,
\newblock ``Adam: A method for stochastic optimization,''
\newblock {\em arXiv preprint arXiv:1412.6980}, 2014.

\bibitem{xue2013gradient}
Wufeng Xue, Lei Zhang, Xuanqin Mou, and Alan~C Bovik,
\newblock ``Gradient magnitude similarity deviation: A highly efficient
  perceptual image quality index,''
\newblock {\em IEEE Transactions on Image Processing}, vol. 23, no. 2, pp.
  684--695, 2013.

\bibitem{bjontegaard2001calculation}
Gisle Bjontegaard,
\newblock ``Calculation of average {PSNR} differences between {RD}-curves,''
\newblock {\em VCEG-M33}, 2001.

\end{thebibliography}

\end{document}